\documentstyle[epsbox,12pt]{article}
\newcommand{\simgt}{\lower.5ex\hbox{$\; \buildrel > \over \sim \;$}}
\newcommand{\simlt}{\lower.5ex\hbox{$\; \buildrel < \over \sim \;$}}
\begin{document}
\title{Thermodynamic properties of nuclear ``pasta'' in neutron star crusts}
\author{Gentaro Watanabe$^{\rm a}$, Kei Iida$^{\rm a,b}$, 
Katsuhiko Sato$^{\rm a,c}$\\
{\it $^{\rm a}$Department of Physics, University of Tokyo,
7-3-1 Hongo, Bunkyo,}\\ {\it Tokyo 113-0033, Japan}\\
{\it $^{\rm b}$Department of Physics, University of Illinois
at Urbana-Champaign,}\\ 
{\it 1110 West Green Street, Urbana, IL 61801-3080, USA}\\
{\it $^{\rm c}$Research Center for the Early Universe, 
University of Tokyo,}\\
{\it 7-3-1 Hongo, Bunkyo, Tokyo 113-0033, Japan}}
\maketitle

\baselineskip=24pt

\begin{abstract}

Equilibrium phase diagrams for neutron star matter at subnuclear
densities are obtained at zero temperature.
Spherical, rod-like and slab-like
nuclei as well as spherical and rod-like nuclear bubbles
are taken into account by using a compressible liquid-drop model.
This model is designed to incorporate uncertainties in the nuclear
surface tension and in the proton chemical potential in a gas of
dripped neutrons. The resultant phase diagrams show that
for typical values of these quantities,
the phases with rod-like nuclei and with slab-like nuclei occur 
in the form of Coulomb lattice
at densities below a point where the system becomes uniform.
Thermal fluctuations leading to displacements of such nuclei
from their equilibrium positions
are considered through explicit evaluations of their elastic constants;
these fluctuations can be effective at destroying the 
layered lattice of slab-like nuclei in the temperature region 
typical of matter in the neutron star crust.

\noindent
$PACS:$ 26.60.+c; 97.60.Jd

\noindent
$Keywords:$ Dense matter; Ground state; Thermal fluctuations; Neutron stars

\end{abstract}
\newpage

\section{Introduction}

In the outer part of a neutron star, nuclei are considered to 
mainly determine the state of matter in equilibrium \cite{review}.  
Except for a thin envelope of the star,
due to Coulomb interactions, nuclei form a bcc lattice
neutralized by a roughly uniform sea of electrons.
With increasing density, weak interactions render
nuclei neutron-rich via electron captures.
Then, at a density of about $\sim 4 \times 10^{11}$ g cm$^{-3}$, 
neutrons begin to drip out of these nuclei. 
The crystalline region of the star (not) including a gas of the dripped 
neutrons is usually referred to as an inner (outer) crust.
In the deepest region of the inner crust, corresponding to
densities just below the normal saturation density 
$\rho_{\rm s}\sim 3\times10^{14}$ g cm$^{-3}$, 
not only are nuclei expected to have rod-like and slab-like shapes,
but also the system is expected to turn inside out in such a way 
that the constituents of the original nuclei form a liquid containing
rod-like and roughly spherical bubbles of the dripped neutrons.
These transformations, 
as originally indicated by Ravenhall et al.\ \cite{ravenhall}
and Hashimoto et al.\ \cite{hashimoto}, stem from 
a delicate competition between the nuclear surface and 
Coulomb energies at small internuclear spacings.

Recent calculations of the ground state of matter in the crust, performed 
by using specific nuclear models \cite{lorenz,oyamatsu,sumiyoshi}, 
indicate that at a density of order $10^{13}$ g cm$^{-3}$, rather
small compared with $\rho_{\rm s}$,
roughly spherical nuclei turn into elongated rod-like nuclei.
Accordingly, the lattice structure is changed from
a regular bcc lattice to a two-dimensional triangular lattice.
With further increasing density, 
these rod-like nuclei are transformed into slab-like nuclei, 
which are arranged in the form of a layered lattice. 
After that, a two-dimensional triangular lattice of rod-like 
bubbles and a bcc Coulomb lattice of roughly spherical bubbles
appear in turn.  At last, at a density of about $\rho_{\rm s}/2$,
the system dissolves into uniform nuclear matter.  
Since slabs and rods look like ``lasagna'' and ``spaghetti'',
the phases with positional order of one- and two-dimension
are often referred to as ``pasta'' phases.

It is of interest to note the relevance of non-spherical nuclei
to pulsar glitches and cooling of neutron stars.
Pulsar glitches are usually considered to be manifestation of a  
sudden large-scale release of neutron vortices trapped
by some pinning centers in the crust \cite{ruderman}.
The force needed to pin vortices has yet to be 
clarified completely even for a bcc lattice of spherical nuclei 
due mainly to the uncertain properties of impurities and defects 
which may be involved in the vortex pinning \cite{jones}. 
Whether or not neutron vortex pinning occurs 
in the ``pasta'' phases possibly formed depends also 
on the unknown properties of impurities and defects in the lattice.
The presence of non-spherical nuclei would accelerate the cooling
of the corresponding region of neutron stars by opening
semi-leptonic weak processes which are unlikely to occur for 
spherical nuclei \cite{lorenz}. 
This stems from the fact that in non-spherical
nuclei, protons have a continuous spectrum at the Fermi surface
in the elongated direction.  
However, such additional cooling would be rather small 
compared with the overall cooling of the stars, since the mass of
the region where non-spherical nuclei are expected to occur 
is only a tiny fraction of the total mass of the star.

At what densities the phases with non-spherical nuclei and with bubbles 
are energetically more favorable than the usual bcc phase and 
the phase of uniform nuclear matter depends on the properties of 
neutron-rich nuclei and of the surrounding pure neutron gas.
The quantities that mainly describe such properties but are still uncertain 
are the nuclear surface tension $E_{\rm surf}$ and the energy required to add
a proton to the pure neutron gas, i.e., the proton chemical potential 
$\mu_{\rm p}^{(0)}$.
$E_{\rm surf}$ controls the size of the nuclei and bubbles, and hence
the sum of the electrostatic and interfacial energies.
With increasing $E_{\rm surf}$ and so this energy sum, 
the density $\rho_{\rm m}$ at which 
the system becomes uniform is lowered. 
There is also a tendency that
the lower $\mu_{\rm p}^{(0)}$, the smaller $\rho_{\rm m}$.
This is because $-\mu_{\rm p}^{(0)}$ represents the
degree to which the neutron gas favors the presence of protons in itself.
This feature is implied by the work of
Arponen \cite{arponen} who studied the sensitivity of the properties of 
uniform nuclear matter at low proton fraction
to the melting density $\rho_{\rm m}$.
Although $E_{\rm surf}$ and $\mu_{\rm p}^{(0)}$ have not been fully 
determined, previous authors 
(e.g., Refs.\ \cite{lorenz,oyamatsu,sumiyoshi})
designated the values of these quantities almost uniquely.
In this paper, by generalizing
a compressible liquid-drop model developed by
Baym, Bethe and Pethick \cite{BBP} (hereafter denoted by BBP)
in such a way as to incorporate uncertainties 
in $E_{\rm surf}$ and $\mu_{\rm p}^{(0)}$,
we draw the equilibrium phase diagrams of zero-temperature
neutron star matter at subnuclear densities. 
The resulting phase diagrams show that
while the phases with rod-like bubbles and with spherical
bubbles can exist only for unrealistically low $E_{\rm surf}$, 
the phases with rod-like nuclei and with slab-like nuclei 
survive almost independently of $E_{\rm surf}$ and $\mu_{\rm p}^{(0)}$.

For these two phases, there are directions in which the system 
is translationally invariant.  
As noted by Pethick and Potekhin \cite{potekhin},
this situation is geometrically analogous with a liquid crystal rather 
than with a rigid solid.  Elastic properties of the nuclear rods and slabs 
are thus characterized by elastic constants used for the corresponding
liquid-crystal phases, i.e., columnar phases and 
smectics A, respectively.  They expressed such constants in terms of
the interfacial and Coulomb energies. 
These energies, as well as the energies of bulk nucleon matter and of a 
electron gas, are well described by the zero-temperature approximation.
This is because the temperature of the crust $\simlt10^{9}$ K,
as inferred from X-ray observations of rotation-powered pulsars, is 
small compared with typical nuclear and electronic excitation energies.
However, such temperature may have consequence to
the structure of the Coulomb lattices involved.  
For example, the outer boundary of the crust is acutely 
controlled by the temperature because of the density dependence
of the melting temperature of the bcc lattice \cite{vanhorn}.
In the deepest region of the crust, 
thermal fluctuations may possibly spoil the 
ordering sustained by the layered and triangular lattices,
as expected from the usual Landau-Peierls argument.  
We thus calculate thermally induced displacements of non-spherical nuclei 
by using the elastic constants given by Pethick and Potekhin 
\cite{potekhin}. 
The critical temperatures at which the ordered structure is 
destroyed by the thermal fluctuations are also estimated.

In Section 2 we construct a compressible
liquid-drop model for nuclei and bubbles, and we write down
equilibrium conditions for zero-temperature matter in the crust. 
The resultant phase diagrams are displayed in Section 3.
Properties of the phase transitions involved are discussed in Section 4.
In Section 5 displacements of rod-like and slab-like nuclei 
are calculated at finite temperatures.
Conclusions are given in Section 6.

\section{Model for neutron star matter at subnuclear densities}

In this section, we write the energy of zero-temperature matter 
in the crust and the conditions for its equilibrium.
We consider five phases  
which contain spherical nuclei, cylindrical nuclei, 
planar nuclei, cylindrical bubbles and spherical bubbles, respectively.
Each phase is taken to be composed of a single species of 
nucleus or bubble at a given baryon density $n_{\rm b}$.
Crystalline structures of these phases are taken into account 
in the Wigner-Seitz approximation, 
which leads to sufficiently accurate evaluations 
of the lattice energy \cite{oyamatsu}.  
In this approximation,
a cell in the bcc lattice, including
a spherical nucleus or bubble of radius $r_{\rm N}$, 
is replaced by a Wigner-Seitz cell defined as
a sphere having radius $r_{\rm c}$ and the same center.  
A cylindrical nucleus or bubble having an infinitely long axis 
and a circular section of radius $r_{\rm N}$ is taken to be contained 
in a cylindrical Wigner-Seitz cell having the same axis and 
a circular section of radius $r_{\rm c}$ 
in place of a cell in the two-dimensional triangular lattice.   
For a planar nucleus with width $2r_{\rm N}$,
a cell in the one-dimensional layered lattice, having width 
$2r_{\rm c}$, is identified with the corresponding Wigner-Seitz cell.
The values of $r_{\rm c}$ for these phases are chosen so that each
Wigner-Seitz cell may have zero net charge and hence have the same
volume as that of the original cell.

\subsection{Energy of matter}

For the purpose of constructing a formula for the
energy of the system allowing for uncertainties in $E_{\rm surf}$
and $\mu_{\rm p}^{(0)}$, 
a compressible liquid-drop model for nuclei and bubbles
is of practical use, which is
characterized by the energy densities of uniform 
nuclear matter and neutron matter as well as by the interfacial
and Coulomb energies.
In this model, neutrons and protons are assumed to be distributed 
uniformly inside and outside the nuclei or bubbles. 
A sea of electrons, which ensures 
the charge neutrality of the system, 
is taken to be homogeneous.
We thus write the total energy density $E_{\rm tot}$ as
\begin{equation}
E_{\rm tot}=\left\{
\begin{array}{ll}
w_{\rm N}+w_{\rm L}+(1-u)E_{\rm n}(n_{\rm n})
+E_{\rm e}(n_{\rm e})& \quad \mbox{(nuclei)}\ , \\
w_{\rm N}+w_{\rm L}+uE_{\rm n}(n_{\rm n})
+E_{\rm e}(n_{\rm e})& \quad \mbox{(bubbles)}\ ,
\end{array}\right.
\end{equation}
where 
$w_{\rm N}$ is the energy of the nuclear matter region
in a cell as divided by the cell volume, 
$w_{\rm L}$ is the lattice energy 
per unit volume as given for the finite-size nuclei or bubbles, 
$n_{\rm n}$ is the number density of neutrons outside the nuclei or 
inside the bubbles,
$n_{\rm e}$ is the number density of the electrons,
$E_{\rm n}$ and $E_{\rm e}$ are
the energy densities of the neutron matter and of the electron gas,
respectively, and  
$u$ is the volume fraction occupied by the nuclei or bubbles,
\begin{equation}
u = \left(\frac{r_{\rm N}}{r_{\rm c}} \right)^d = \left\{
\begin{array}{ll}
\displaystyle{\frac{n_{\rm b}-n_{\rm n}}{n-n_{\rm n}}}
& \quad \mbox{(nuclei)\ ,}\smallskip\\
\displaystyle{\frac{n-n_{\rm b}}{n-n_{\rm n}}}
& \quad \mbox{(bubbles)\ ,}
\end{array}\right.
\end{equation}
where $d$ is the dimensionality defined as $d=1$ for slabs, 
$d=2$ for cylinders and $d=3$ for spheres, and
$n$ is the nucleon number density in the nuclear matter region.
In Eq.\ (1), we ignore corrections due to nucleon pairing effects,
shell and curvature effects in the nuclei and bubbles, 
and deformations and surface diffuseness of the nuclei and bubbles.

In writing $w_{\rm N}$, 
we adopt a generalized version of the 
compressible liquid drop model developed by BBP \cite{BBP}.
$w_{\rm N}$ is then expressed as
\begin{equation}
w_{\rm N}(n,x,n_{\rm n},r_{\rm N},r_{\rm c},d) = \left\{
\begin{array}{ll}
un[(1-x)m_{\rm n}+xm_{\rm p}]c^2+unW(k,x)\nonumber\\
\quad
+w_{\rm surf}(n,x,n_{\rm n},r_{\rm N},u,d)
+w_{\rm C}(n,x,r_{\rm N},u,d)
& \quad \mbox{(nuclei)\ ,}\\ \\
(1-u)n[(1-x)m_{\rm n}+xm_{\rm p}]c^2+(1-u)nW(k,x)\nonumber\\
\quad
+w_{\rm surf}(n,x,n_{\rm n},r_{\rm N},u,d)
+w_{\rm C}(n,x,r_{\rm N},u,d)
& \quad \mbox{(bubbles)\ ,}
\end{array}\right.
\end{equation}
where 
$m_{\rm n}$ $(m_{\rm p})$ is the neutron (proton) rest mass,
$W(k,x)$ is the energy per nucleon for uniform nuclear matter
of nucleon Fermi wave number $k=(3\pi^{2}n/2)^{1/3}$ and proton 
fraction $x$, as given by BBP [see Eq.\ (3.19) in Ref.\ \cite{BBP}],
$w_{\rm surf}$ is the nuclear surface energy per unit volume,
and $w_{\rm C}$ is the self Coulomb energy (per unit volume) 
of protons contained in a cell.

$W(k,x)$ includes four parameters characterizing the saturation
properties of nearly symmetric nuclear matter,
which are the binding energy $w_0$, 
the saturation density $2k_0^{3}/3\pi^{2}$,
the incompressibility $K$ and the coefficient $s$ determining the
symmetry energy.
In the present work, the values for these parameters determined
empirically by BBP in such a way as to reproduce nuclear masses and 
radii of $\beta$ stable nuclei are replaced by the following values
that are generally accepted among recent literature: 
$w_0=16.0$ MeV, $k_0=1.36$ fm$^{-1}$, $K=230$ MeV and $s=30.0$ MeV.
We remark in passing that such a replacement makes
no significant difference in the phase diagrams for neutron star matter
at subnuclear densities that will be exhibited in Section 3.

The proton chemical potential $\mu_{\rm p}^{(0)}$ in pure neutron matter
is contained in $W(k,x)$.  This quantity,
dominating stability of the phases including a dripped neutron gas
over the phase of uniform nuclear matter, has not been well determined 
from microscopic calculations based on nucleon-nucleon interactions.
We thus set $\mu_{\rm p}^{(0)}$ (not including the rest mass) as 
\begin{equation}
\mu_{\rm p}^{(0)}=-C_1 n_{\rm n}^{2/3}\ ,
\end{equation}
where $C_{1}$ is the free parameter being positive definite.
As can be seen from Fig.\ 1,
Eq.\ (4) approximately reproduces the overall density dependence of the
results obtained from the Hartree-Fock theory with Skyrme interactions
(see Ref.\ \cite{pethick} and references therein) and from 
the lowest-order Brueckner theory with the Reid soft-core potential
\cite{siemens,sj}.
Hereafter, $C_{1}$ will be set as $C_{1}=300,400,500,600$ MeV fm$^{2}$; 
the case of $C_{1}=300$ MeV fm$^{2}$ agrees well with the result
of Siemens and Pandharipande \cite{siemens} adopted by BBP,
whereas the case of $C_{1}=400$ MeV fm$^{2}$ is consistent
with the other results cited in Fig.\ 1 for densities of
interest to us $n\simlt0.12$ fm$^{-3}$.

In constructing the surface energy per unit volume $w_{\rm surf}$,
we take note of the expression given by BBP [see Eq.\ (4.30) in Ref.\ 
\cite{BBP}] and the Hartree-Fock
calculations by Ravenhall, Bennett and Pethick \cite{RBP} (hereafter
denoted by RBP) using a Skyrme interaction.
For comparison, it is useful to consider the surface tension
$E_{\rm surf}$ that is independent of the shapes of nuclei and bubbles; 
this is related to $w_{\rm surf}$ via
\begin{equation}
E_{\rm surf} = \frac{r_{\rm N}w_{\rm surf}}{ud}\ .
\end{equation}
Following the spirit of Iida and Sato \cite{iida},
we express $E_{\rm surf}$ as  
\begin{equation}
E_{\rm surf}=C_{2}\tanh\left(\frac{C_{3}}{\mu_{\rm n}^{(0)}}\right) 
E_{\rm surf}^{\rm BBP}\ ,
\end{equation}
where 
$C_{2}$ and $C_{3}$ are the adjustable parameters
that will be discussed just below, 
$\mu_{\rm n}^{(0)}=\partial E_{\rm n}/\partial n_{\rm n} - m_{\rm n}c^2$ 
is the neutron chemical potential in the neutron gas 
not including the rest mass, and 
\begin{equation}
E_{\rm surf}^{\rm BBP}=\frac{\sigma(n-n_{\rm n})^{2/3}[W(k_{n},0)-W(k,x)]}
{(36\pi)^{1/3}w_{0}}\ ,
\end{equation}
with $k_{n}=(3\pi^{2}n_{n}/2)^{1/3}$ and $\sigma=21.0$ MeV.
If we set $C_{2}=1.0$ and $C_{3}=3.5$ MeV for $C_{1}=400$ MeV fm$^{2}$,
as shown in Fig.\ 2,
the surface tension calculated for matter in equilibrium 
(see Subsection 2.2 for details) agrees well with
the RBP values for proton fractions of interest here $x\simlt0.2$.
For $C_{1}=300,500,600$ MeV fm$^{2}$, consistency with the RBP values
is also obtained from $C_{2}=1.0$ and $C_{3}=3.5$ MeV.
Hereafter this value of $C_{3}$ will be fixed, whereas, 
for our present purpose of examining the dependence on nuclear models
of the equilibrium phase diagrams for matter in the crust, 
we shall shake the values of $C_{2}$ within a rather wide range covering 
$C_{2}=1.0$ (see Section 3).

We proceed to express the total electrostatic energy per unit volume
$w_{\rm C+L}=w_{\rm C}+w_{\rm L}$.  In the Wigner-Seitz approximation
adopted here, it reads \cite{review}
\begin{equation}
w_{\rm C+L}=2\pi(exn r_{\rm N})^2 u f_d(u)\ ,
\end{equation}
with
\begin{equation}
f_d(u)= \frac{1}{d+2}
\left[\frac{2}{d-2} \left( 1-\frac{d u^{1-2/d}}{2} \right) +u \right]\ .
\end{equation}

The energy density of the pure neutron gas, 
whose density derivative $\mu_{\rm n}^{(0)}$, 
together with $E_{\rm surf}$ and $\mu_{\rm p}^{(0)}$, plays a role 
in determining the melting density $\rho_{\rm m}$ \cite{pethick}, 
is set by using $W(k,x)$ as
\begin{equation}
E_{\rm n}(n_{\rm n})=[W(k_{\rm n},0)+m_{\rm n}c^2]n_{\rm n}\ .
\end{equation}
Note that this energy density corresponds to the fitting formula 
determined by BBP [see Eq.\ (3.8) in Ref.\ \cite{BBP}]
so as to reproduce the data for $k_{\rm n}\simlt1.7$ fm$^{-3}$
obtained by Siemens and Pandharipande \cite{siemens}
in the lowest-order Brueckner theory with the Reid soft-core potential. 
These data hold good at least for $k_{\rm n}\simlt k_{0}$,
as can be seen from comparison with the results from several works based on 
more elaborate many-body schemes and potential models \cite{pethick}.
We shall thus use the BBP formula without any modification.

The energy density of the free electron gas, which is relativistic and 
degenerate, is finally
expressed as
\begin{equation}
E_{\rm e}(n_{\rm e})=\frac{3}{4} \hbar c k_{\rm e}n_{\rm e}\ ,
\end{equation}
with the electron Fermi wave number $k_{\rm e}=(3\pi^{2}n_{\rm e})^{1/3}$.
Here, we ignore the electron mass and the electron Hartree-Fock energy.

\subsection{Equilibrium conditions}

As a next step towards the description of the equilibrium phase diagrams
for crustal matter, we minimize the energy density $E_{\rm tot}$ with
respect to five variables $n$, $x$, $n_{\rm n}$, $r_{\rm N}$ and $u$
under fixed baryon density $n_{\rm b}$,
\begin{equation}
n_{\rm b} = \left\{
\begin{array}{ll}
un+(1-u)n_{\rm n} & \quad \mbox{(nuclei)\ ,}\\
(1-u)n+un_{\rm n} & \quad \mbox{(bubbles)\ .}
\end{array}\right.
\end{equation}
and under charge neutrality,
\begin{equation}
n_{\rm e} = \left\{
\begin{array}{ll}
xnu & \quad \mbox{(nuclei)\ ,}\\
xn(1-u) & \quad \mbox{(bubbles)\ .}
\end{array}\right.
\end{equation}
Such minimization results in four equilibrium conditions,
which were given by BBP for spherical nuclei 
(see Section 2 in Ref.\ \cite{BBP}).

Optimization of $E_{\rm tot}$ with respect to $r_{N}$
at fixed $n$, $x$, $n_{\rm n}$ and $u$ leads to
a familiar relation for size-equilibrium,
\begin{equation}
w_{\rm surf}=2w_{\rm C+L}\ .
\end{equation}
This relation is thus independent of dimensionality $d$. 
Such independence was understood \cite{review}
from the fact that the system,  no matter what type of nucleus or bubble
it may contain, is in a three-dimensional physical space.

The condition for equilibrium of the dripped neutrons with the neutrons
in the nuclear matter region, arising from minimization of $E_{\rm tot}$
with respect to $n$ at fixed $n_{\rm b}$, $nx$, $r_{\rm N}$ and $u$,
reads
\begin{equation}
\mu_{\rm n}^{\rm (N)}=\mu_{\rm n}^{\rm (G)}\ .
\end{equation}
Here
\begin{equation}
\mu_{\rm n}^{\rm (N)} = \left\{
\begin{array}{ll}
\displaystyle{\left. \frac{\partial (w_{\rm N}+w_{\rm L})}
{u\partial n}\right|
_{nx, n_{\rm n}, r_{\rm N}, u} - m_{\rm n}c^2}
& \quad \mbox{(nuclei)}\\ \\
\displaystyle{\left. \frac{1}{1-u}
\frac{\partial (w_{\rm N}+w_{\rm L})}
{\partial n}\right|
_{nx, n_{\rm n}, r_{\rm N}, u} - m_{\rm n}c^2}
& \quad \mbox{(bubbles)}
\end{array}\right.
\end{equation}
is the neutron chemical potential in the nuclear matter region, and 
\begin{equation}
\mu_{\rm n}^{\rm (G)} = \left\{
\begin{array}{ll}
\displaystyle{
\left( \frac{\partial E_{\rm n}}{\partial n_{\rm n}} - m_{\rm n}c^2 \right)
+ \left.\frac{1}{1-u}
\frac{\partial w_{\rm N}}{\partial n_{\rm n}} \right|
_{n, x, r_{\rm N}, u}}
& \quad \mbox{(nuclei)}\\ \\
\displaystyle{
\left( \frac{\partial E_{\rm n}}{\partial n_{\rm n}} - m_{\rm n}c^2 \right)
+ \left.\frac{1}{u}
\frac{\partial w_{\rm N}}{\partial n_{\rm n}} \right|
_{n, x, r_{\rm N}, u}}
& \quad \mbox{(bubbles)}
\end{array}\right.
\end{equation}
is the neutron chemical potential in the neutron gas 
modified by the surface energy.

The condition that the nuclear matter region is stable against
$\beta$-decay is yielded by minimization of $E_{\rm tot}$
with respect to $x$ at fixed $n$, $n_{\rm n}$, $r_{\rm N}$ and $u$.
It is expressed as
\begin{equation}
\mu_{e}-(m_{\rm n}-m_{\rm p})c^{2}=
\mu_{\rm n}^{\rm (N)}-\mu_{\rm p}^{\rm (N)}\ ,
\end{equation}
where 
\begin{equation}
\mu_{e}=\hbar k_{e}c
\end{equation}
is the electron chemical potential, and 
\begin{equation}
\mu_{\rm p}^{\rm (N)}= \left\{
\begin{array}{ll}
\displaystyle{\left.\frac{\partial (w_{\rm N}+w_{\rm L})}
{u\partial (nx)}\right|_{n(1-x), n_{\rm n}, r_{\rm N}, u} 
- m_{\rm p}c^2}
& \quad \mbox{(nuclei)}\\ \\
\displaystyle{\left.\frac{1}{1-u}
\frac{\partial (w_{\rm N}+w_{\rm L})}{\partial (nx)}
\right|_{n(1-x), n_{\rm n}, r_{\rm N}, u} - m_{\rm p}c^2}
& \quad \mbox{(bubbles)}
\end{array}\right.
\end{equation}
is the proton chemical potential in the nuclear matter region.

Finally, minimizing $E_{\rm tot}$ with respect to $r_{\rm N}$ 
at fixed $un$, $x$, $r_{\rm c}$ and $(1-u)n_{\rm n}$ for nuclei
[$(1-u)n$, $x$, $r_{\rm c}$ and $un_{\rm n}$ for bubbles]
yields the condition for pressure equilibrium 
between the nuclear matter and the neutron gas,
\begin{equation}
P^{({\rm N})}=P^{({\rm G})}\ ,
\end{equation}
where
\begin{equation}
P^{({\rm N})}=\left\{
\begin{array}{ll}
\displaystyle{\left.
-\frac{\partial (w_{\rm N}+w_{\rm L})}{\partial u}
\right|_{nu, x, n_{\rm n}, r_{\rm c}}}
& \quad \mbox{(nuclei)}\\ \\
\displaystyle{\left.
\frac{\partial (w_{\rm N}+w_{\rm L})}{\partial u}
\right|_{n(1-u), x, n_{\rm n}, r_{\rm c}}}
& \quad \mbox{(bubbles)}
\end{array}\right.
\end{equation}
is the pressure of the nuclear matter region, and 
\begin{equation}
P^{({\rm G})}=n_{\rm n}[\mu_{\rm n}^{({\rm G})} - W(k_{\rm n},0)]
\end{equation}
is the pressure of the neutron gas.

At given $n_{\rm b}$ and subject to charge neutrality (13), 
we have calculated the total energy density $E_{\rm tot}$ 
for the five phases containing spherical nuclei, cylindrical nuclei, 
planar nuclei, spherical bubbles and cylindrical bubbles, respectively.
In such calculations, we first express $x$, $r_{\rm N}$ 
and $r_{\rm c}$ in terms of $n_{\rm b}$, $n$ and $n_{\rm n}$
by using conditions (12), (14) and (18).
In this stage, we can rewrite 
$\mu_{\rm n}^{\rm (N)}$, $\mu_{\rm n}^{\rm (G)}$ and $P^{\rm (N)}$
given by Eqs.\ (16), (17) and (22) as 
\begin{equation}
\mu_{\rm n}^{\rm (N)}= 
 \left\{
\begin{array}{ll}
\displaystyle{W(k,x)+\frac{k}{3}\frac{\partial W(k,x)}{\partial k}
+ \frac{d}{r_{\rm N}}
\left(\frac{1}{n}
+ \frac{\partial}{\partial n}\right)E_{\rm surf}
+ x[\mu_{\rm e}-(m_{\rm n}-m_{\rm p})c^2]}
& \mbox{(nuclei)\ ,}\smallskip\\
\displaystyle{W(k,x)+\frac{k}{3} \frac{\partial W(k,x)}{\partial k}
+ \frac{d}{r_{\rm N}}
\left(\frac{u}{1-u}\right) 
\left(\frac{1}{n}
+ \frac{\partial}{\partial n}\right)E_{\rm surf}
+ x[\mu_{\rm e}-(m_{\rm n}-m_{\rm p})c^2]}
& \mbox{(bubbles)\ ,}
\end{array}\right.
\end{equation}
\begin{equation}
\mu_{\rm n}^{\rm (G)}= \left\{
\begin{array}{ll}
\displaystyle{W(k_{\rm n},0)+\frac{k_{\rm n}}{3}
\frac{\partial W(k_{\rm n},0)}{\partial k_{\rm n}}
+ \frac{d}{r_{\rm N}} \frac{u}{1-u} 
\frac{\partial E_{\rm surf}}{\partial n_{\rm n}}}
& \mbox{(nuclei)\ ,}\smallskip\\
\displaystyle{W(k_{\rm n},0)+\frac{k_{\rm n}}{3}
\frac{\partial W(k_{\rm n},0)}{\partial k_{\rm n}}
+ \frac{d}{r_{\rm N}} 
\frac{\partial E_{\rm surf}}{\partial n_{\rm n}}}
& \mbox{(bubbles)\ ,}
\end{array}\right.
\end{equation}
\begin{equation}
P^{({\rm N})}= \left\{
\begin{array}{ll} 
\displaystyle{\frac{nk}{3} \frac{\partial W(k,x)}{\partial k}
- \frac{d-1}{r_{\rm N}} E_{\rm surf}
+ \frac{dn}{r_{\rm N}} \frac{\partial E_{\rm surf}}{\partial n}
+ \frac{4\pi r_{\rm N}^2}{d(d+2)} (exn)^2 (1-u)}
& \mbox{(nuclei)\ ,}\smallskip\\
\displaystyle{\frac{nk}{3} \frac{\partial W(k,x)}{\partial k}
+ \frac{d-1}{r_{\rm N}} E_{\rm surf}
+ \frac{dn}{r_{\rm N}} \frac{u}{1-u} \frac{\partial E_{\rm surf}}{\partial n}
- \frac{4\pi r_{\rm N}^2}{d(d+2)} (exn)^2 (1-u)}
& \mbox{(bubbles)\ .}
\end{array}\right.
\end{equation}
Conditions (15) and (21) in turn give $n$ and $n_{\rm n}$
as function of $n_{\rm b}$.  
Substitution of these values into Eq.\ (1)
leads thus to the minimum values of $E_{\rm tot}$ 
at given $n_{\rm b}$ and conformation.
The energy density for $\beta$-equilibrated uniform nuclear matter
has also been evaluated at the same $n_{\rm b}$
by using the energy densities $nW(k,x)$ and $E_{\rm e}(n_{\rm e})$
and by following a line of argument of BBP 
(see Section 8 in Ref.\ \cite{BBP}). 
Here we ignore the presence of muons, the number density of which
is, if any, far smaller than $n_{e}$ for densities of interest here.
We have finally found the phase giving the smallest energy density among
the uniform and crystalline phases; the resultant
phase diagrams for the ground-state matter
at subnuclear densities will be shown in the next section.

\section{Phase diagrams}

We may now draw the phase diagrams for neutron star matter in 
the ground state on the $n_{\rm b}$ versus $C_{2}$ plane
for $C_{1}=300,400,500,600$ MeV fm$^{2}$.\footnote{Note that 
these diagrams are different from the usual ones written 
over the density versus temperature plane.}
The results have been plotted in Fig.\ 3.
We have thus confirmed the tendency, as mentioned in Section 1,
that larger $E_{\rm surf}$ and lower $\mu_{\rm p}^{(0)}$ 
help the uniform matter phase ``erode'' the phases 
with non-spherical nuclei and with bubbles.

As can be found from Fig.\ 3, the obtained phase diagrams
basically reproduce the feature that with increasing density, 
the shape of the nuclear matter region changes like 
sphere $\rightarrow$
cylinder $\rightarrow$ slab $\rightarrow$ cylindrical hole
$\rightarrow$ spherical hole $\rightarrow$ uniform matter.
This feature can be derived from simple discussions 
about the Coulomb and surface effects 
ignoring the bulk energy \cite{hashimoto}.
We also find that the energy difference between two successive phases
is generally of order 0.1--1 keV per nucleon (see Fig.\ 4);
this result is consistent with that obtained by others in earlier 
investigations (e.g., Refs.\ \cite{lorenz,oyamatsu,sumiyoshi}).
However, it is significant to notice that 
for most sets of $C_{1}$ and $C_{2}$ including
the typical one ($C_{1}=400$ MeV fm$^{2}$ and $C_{2}=1.0$), 
the shape changes become simpler, i.e., 
sphere $\rightarrow$ cylinder $\rightarrow$ slab
$\rightarrow$ uniform matter.
This behavior, 
due probably to bulk corrections to the Coulomb and surface effects,
was not observed in the previous works
(see, e.g., Refs.\ \cite{lorenz,oyamatsu,sumiyoshi}).
Since the values of $E_{\rm n}$, $\mu_{\rm p}^{(0)}$ and $E_{\rm surf}$
as adopted by these works are consistent with those obtained here 
for $C_{1}=400$ MeV fm$^{2}$ and $C_{2}=1.0$, we may conclude that 
the difference in the phase diagrams should originate mainly from that 
in the properties of uniform nuclear matter 
at low but $finite$ proton fractions, i.e., $x\sim0.1$.  
These properties depend on the adopted schemes for constructing interpolation
between almost pure neutron matter and nearly symmetric nuclear matter 
in the absence of reliable microscopic calculations of the energy density 
of uniform nuclear matter at intermediate proton fractions. 
Better understanding of the equilibrium properties of 
neutron star matter at subnuclear densities
will thus require future theoretical and/or experimental works 
covering the region of intermediate proton fractions.

Curvature effects in nuclei and bubbles, which were at least
implicitly taken into account in Refs.\ \cite{lorenz,oyamatsu,sumiyoshi}
while have been ignored in the present analysis,
may be another source of the difference in the phase diagrams 
from such earlier works.
This is due to the conformation dependence of these effects:
$w_{\rm curv}$(sphere)$>w_{\rm curv}$(cylinder)
$>w_{\rm curv}$(slab)
$=0>w_{\rm curv}$(cylindrical hole)
$>w_{\rm curv}$(spherical hole),
where $w_{\rm curv}$ is the curvature energy per unit volume.
The degree of the resultant stability of the higher-density 
crystalline phase over the lower-density one depends on the 
values of $w_{\rm curv}$, which are also uncertain at $x\sim0.1$.
If we determine these values from the curvature thermodynamic 
potential obtained by Kolehmainen et al.\ \cite{kolehmainen} in the 
Thomas-Fermi theory with Skyrme interactions,
the energy gain of the higher-density crystalline phase with 
respect to the lower-density one amounts to a few keV per nucleon.
For our typical nuclear model ($C_1=400$ MeV fm$^2$ and $C_2=1.0$),
as can be seen from Fig.\ 4, such curvature effects do not lead to
appearance of the density region where the phases with bubbles 
are stable, but do lower the transition density from sphere to 
cylinder and that from cylinder to slab by 10\% or so.
In Fig.\ 3, correspondingly, the region surrounded by the bcc phase and
the uniform phase is expected to enlarge.

\section{Properties of phase transitions}

The structural transitions examined in the previous section
are first-order transitions.
This is because
the above-mentioned sequence of geometrical structure 
is mainly determined by 
the competition between the surface and Coulomb energies.
In order to confirm this property, 
we have plotted in Fig.\ 5
the sizes of the nucleus or bubble and of the Wigner-Seitz
cell, $r_{\rm N}$ and $r_{\rm c}$, evaluated as functions
of $n_{\rm b}$ for $C_{1}=400$ MeV fm$^{2}$ and $C_{2}=1.0$.
It can be clearly seen from this figure that
$r_{\rm N}$ and $r_{\rm c}$ jump at the transition points. 
It is of interest to note that for $C_{2}=0.01$, in which all the types 
of crystalline phases occur, the cell size $r_{\rm c}$ shows
a dependence on the dimensionality $d$ such that it 
is largest for $d=3$ (spheres and spherical holes) and  
smallest for $d=1$ (slabs).  
This behavior was also obtained from the previous
liquid-drop-model calculations \cite{lorenz}.

The neutron density profile is
useful in probing how the system dissolves into uniform matter.
In Fig.\ 6 we have thus plotted
the neutron densities in the nuclear matter region and the neutron gas
region for the crystalline phases as well as the one 
for the uniform phase, evaluated
as functions of $n_{\rm b}$
for $C_{1}=300,400,500,600$ MeV fm$^{2}$ and $C_{2}=1.0$. 
For these parameter sets, as shown in Fig.\ 3,
there is a melting transition from the phase with slab-like nuclei 
to the uniform phase.
We can observe from Fig.\ 6 that as the density approaches 
the melting point, 
the neutron distribution becomes more and more smooth.
This behavior, as we have confirmed for various values of $C_2$,
is consistent with the results
from the Thomas-Fermi calculations \cite{oyamatsu,ogasawara}.
We also find that at the melting point,
the neutron density profile is discontinuous between the crystalline 
and uniform phases, indicating that the transition is of first order.
However, this discontinuity should be fairly small as compared with  
the case of matter created in stellar collapse.  In this case,   
due to larger proton fraction $x$ yielded by degenerate neutrinos, 
such a discontinuity should become so large that 
neighbouring nuclei (or bubbles) appear to touch and fuse 
with each other at the melting point. 
Detailed estimates made for this material will be 
reported elsewhere.

\section{Thermal fluctuations}

Let us now estimate thermally induced displacements of rod-like
and slab-like nuclei from their equilibrium positions; 
the presence of these nuclei is energetically favored 
for typical values of $C_{1}$ and $C_{2}$, as shown in Section 3.
For such estimates,
we first write the elastic constants for the phases
containing these nuclei by following Pethick and Potekhin 
\cite{potekhin}.
Since the phase with slab-like (rod-like) nuclei
is structurally similar to 
a smectic A (columnar phase) liquid crystal,
they derived the expressions for the elastic constants 
by comparing the energy increase due to deformation, obtained from 
the liquid-drop model for the nuclei in the incompressible limit,
with that used for the corresponding liquid crystals.
For the layered phase composed of slab-like nuclei,
the energy density due to displacement $v$ of layers in their
normal direction (taken to be parallel to the $z$-axis)
can be written as \cite{degennes}
\begin{equation}
F=\frac{B}{2} \left[
\frac{\partial v}{\partial z}-\frac{1}{2}(\nabla_\bot v)^2 \right]^2
+\frac{K_1}{2}(\nabla_\bot^2 v)^2\ .
\end{equation}
Here, the elastic constants $B$ and $K_1$ are expressed as \cite{potekhin}
\begin{eqnarray}
B &=& 6w_{\rm C+L}\ ,\\
K_1 &=& \displaystyle{\frac{2}{15}w_{\rm C+L}(1+2u-2u^2)r_{\rm c}^2}\ .
\end{eqnarray}
For the two-dimensional triangular lattice of rod-like nuclei,
the energy density due to a two-dimensional displacement vector
$\mbox{\boldmath $v$}=(v_{x}, v_{y})$ of cylinders running along 
the $z$-axis is given by \cite{degennes}
\begin{eqnarray}
F &=& \frac{B}{2}
\left(\frac{\partial v_x}{\partial x}+\frac{\partial v_y}{\partial y}\right)^2
+\frac{C}{2}\left[
\left(\frac{\partial v_x}{\partial x}-\frac{\partial v_y}{\partial y}\right)^2
+\left(\frac{\partial v_x}{\partial y}+\frac{\partial v_y}{\partial x}\right)^2
\right] \nonumber\\
&&+ \frac{K_3}{2}
\left(\frac{\partial^2 \mbox{\boldmath $v$}}{\partial z^2}\right)^2
+B'\left(\frac{\partial v_x}{\partial x}+\frac{\partial v_y}{\partial y}\right)
\left(\frac{\partial \mbox{\boldmath $v$}}{\partial z}\right)^2
+\frac{B''}{2} \left(\frac{\partial 
\mbox{\boldmath $v$}}{\partial z}\right)^4\ .
\end{eqnarray}
Here, the elastic constants $B$, $B'$, $B''$, $C$ and $K_3$
read \cite{potekhin}
\begin{eqnarray}
B &=& \displaystyle{\frac{3}{2} w_{\rm C+L}}\ ,\\
B' &=& - \displaystyle{\frac{3}{4} w_{\rm C+L}}\ ,\\
B'' &=& \displaystyle{\frac{3}{8} w_{\rm C+L}}\ ,\\
C &\simeq& 10^{2.1(u-0.3)} w_{\rm C+L}\ ,\\
K_3 &\simeq& 0.0655 w_{\rm C+L}\ r_{\rm c}^2 \quad
\mbox{(for 0.15\simlt $u$\simlt 0.55)}\ .
\end{eqnarray}
In Eqs.\ (28), (29) and (31)--(35), $w_{\rm C+L}$, 
as given by Eq.\ (8), satisfies the size-equilibrium condition (14).

We may then calculate the mean-square displacement at finite
temperature $T$, defined as $\langle |v|^2 \rangle$,
where $\langle\cdots\rangle$ is the
average over the probability distribution proportional to 
$\exp(-\int_{V}dV F/k_{\rm B}T)$ with the lattice volume 
$V$ and the Boltzmann constant $k_{\rm B}$.\footnote{Here 
the fluctuations are treated thermodynamically.
At $T=0$, however, the fluctuations behave quantum-mechanically; 
the resultant mean-square displacements are of order 
$\hbar c_{2}/(r_{\rm c}^{2}w_{\rm C+L})$, where
$c_{2}$ is the velocity of second sound associated with 
the varying inter-nuclear distance \cite{chandra}, roughly
estimated as $c_{2}\sim\sqrt{w_{\rm C+L}/\rho_{\rm m}}$. 
For typical values $r_{\rm c}\sim 10$ fm, $w_{\rm C+L}\sim 10^{-3}$
MeV fm$^{-3}$ and $\rho_{\rm m}\sim10^{14}$ g cm$^{-3}$,
comparison of the quantum fluctuations thus estimated
with the thermal fluctuations to be obtained below
gives rise to the quantum-classical crossover temperature $T_{0}$
of order $10^{7}$ K for planar nuclei and of order $10^{9}$ K 
for cylindrical nuclei.  
Consequently, the forthcoming arguments hold good
for $T\simgt T_{0}$ but underestimate the displacement
for $T\simlt T_{0}$.}
Let us assume that the length scale $L$ of the lattice
is far larger than $r_{\rm c}$.
For the one-dimensional layered lattice,
the mean-square displacement can then be evaluated 
within the harmonic approximation allowing for the terms
up to $O(v^{2})$ in Eq.\ (27) as \cite{chandra}
\begin{equation}
\langle |v|^2 \rangle \simeq \frac{k_{\rm B} T}{4\pi \sqrt{BK_1}}
\ln{\left( \frac{L}{a} \right)}\ ,
\end{equation}
where $a=2r_{\rm c}$ is the layer spacing.
For the two-dimensional triangular lattice,
by retaining the terms up to $O(v_{x}^{2})$ and $O(v_{y}^{2})$
in Eq.\ (30), we obtain \cite{chandra}
\begin{equation}
\langle 
|\mbox{\boldmath $v$}|^{2}
\rangle
\simeq \frac{k_{\rm B} T}{(B+2C) \sqrt{\pi\lambda a}}\ ,
\end{equation}
where 
$\lambda = \sqrt{2K_3/(B+2C)}$, and
$a=(2\pi /\sqrt{3})^{1/2} r_{\rm c}$ is the lattice constant
for hexagonal cells.\footnote{Strictly speaking,
Eq.\ (37) is valid when the length of the nuclei is large compared with
the linear dimension of the lattice in the $xy$ plane.  This situation 
is energetically preferred over the opposite case, in which there is
a larger total sectional area of the nuclei 
at the boundary of the lattice.}
Expressions (36) and (37) are analogous to those appearing
in the context of Landau-Peierls instabilities, 
which destroy a one-dimensional 
ordering in an $infinite$ three-dimensional system.

In Fig.\ 7 (8) we have plotted the ratio 
$\sqrt{\langle |\mbox{\boldmath $v$}|^{2} \rangle}/(a/2 - r_{\rm N})$
of the root-mean-square displacement of the slab-like (rod-like)
nucleus, calculated from Eq.\ (36) [(37)] 
for $k_{\rm B}T=0.1$ MeV, to the 
shortest distance between the surface of the nucleus
in its equilibrium position and
the boundary of the cell containing it. 
In these calculations we have set $L=10\ {\rm km}$, 
which is the typical neutron star radius.
The results for the layered phase, depending on $L$ logarithmically,
are essentially unchanged by the choice of $L$; 
when we set $L=1\ {\rm \mu m}$, for example,
the values shown in Fig.\ 7 are 
only multiplied by a factor of $\simeq0.7$. 
It is obvious from comparison between Figs.\ 7 and 8 that for the same
$C_{1}$ and $C_{2}$ the relative displacement of the slab-like nucleus 
is appreciably large compared with that of the rod-like nucleus,
a feature yielded by the logarithmic factor appearing in Eq.\ (36).
We also find that 
the relative displacement for both lattices decreases
with increasing surface tension (or $C_2$).
This is because the elastic constants, acting to reduce 
the displacements given by Eqs.\ (36) and (37), are 
proportional to the equilibrium value of $w_{\rm C+L}$ 
satisfying condition (14).

It is noteworthy that for the typical parameter set 
($C_{1}=400$ MeV fm$^{2}$ and $C_{2}=1.0$), the relative
displacement for the layered phase at $k_{\rm B}T=0.1$ MeV, 
as plotted in Fig.\ 7, 
takes on values of up to about unity.
The planar nucleus deformed to such an extent
touches the boundary of the neighbouring cell as
defined in the case in which the lattice is in its equilibrium.
This suggests the possibility that the fluctuational displacements
destroy the ordered configuration in a length scale of down to 
$\sim r_{\rm c}$ even at temperatures typical of matter in the neutron 
star crust.\footnote{  
Reconnection of the planar nuclei that may be 
involved in such destruction leads inevitably to departures
from thermodynamic equilibrium, 
which are beyond the scope of this paper.}
In theoretically predicting the interiors of observed 
neutron stars having various temperature profiles, therefore, 
it is useful to estimate as a function of $n_{\rm b}$
the critical temperature
$T_{\rm c}$ at which the relative displacements become unity.
The results obtained for the layered lattice and the triangular
lattice have been plotted in Figs.\ 9 and 10, 
respectively.\footnote{
Note that the values of $T_{\rm c}$ are meaningless
when they are smaller than the crossover temperature $T_{0}$ 
from the classical to the quantum fluctuations; in this case,
quantum effects keep the relative displacement large compared
with unity for $T<T_{0}$.} 
The difference in $T_{\rm c}$ between these two lattices,
as can be observed from Figs.\ 9 and 10, suggests that if
formation of these two lattices can occur dynamically in the star,
the layered phase is formed later than the triangular phase
during the star' s cooling.

\section{Conclusion}

We have examined the dependence on the surface tension $E_{\rm surf}$
and on the proton chemical potential $\mu_{\rm p}^{(0)}$
in pure neutron matter, of the density region in which
the presence of non-spherical nuclei and of bubbles
is energetically favored at $T=0$. 
We have found that as $E_{\rm surf}$ decreases or 
$\mu_{\rm p}^{(0)}$ increases, such a density region becomes larger.
For the values of $E_{\rm surf}$ and $\mu_{\rm p}^{(0)}$
as adopted in recent literature, our results show that 
in the ground state, the phases with rod-like nuclei and with 
slab-like nuclei lie between the bcc lattice phase and 
the uniform nuclear matter phase.
The fluctuational displacements of such non-spherical nuclei
from their equilibrium positions
have been estimated at finite temperature. 
It has been suggested that at temperatures typical of matter 
in the neutron star crust, such fluctuations may
melt the layered lattice of slab-like nuclei.

Even if rod-like and slab-like nuclei are thermodynamically stable, 
whether or not they are actually present 
in the star depends on occurrence of the 
dynamical processes leading to their formation.
These processes are thought to involve
instabilities against nuclear deformation and 
against proton clustering in uniform nuclear matter \cite{review}.
If the occurrence of such processes were confirmed, 
it would become still more significant to
consider effects of the presence of non-spherical nuclei
on the structure and evolution of neutron stars. 
In the context of pulsar glitches, the question of 
what kind of impurities and defects are formed in the lattices of 
non-spherical nuclei would be essential to understanding of the 
mechanism for neutron vortex pinning.
For neutron star cooling, 
it would be interesting to consider the direct URCA processes
in non-spherical nuclei; as suggested by Lorenz et al.\ \cite{lorenz},
these processes might be allowed
by the proton continuous spectrum in the elongated direction
of the nuclei,  in contrast to the case of roughly spherical nuclei, 
and by the band structure of neutrons moving in the
periodic potential created by the nuclei.

\section*{Acknowledgements}

We are grateful to Professor Takeo Izuyama for useful discussion
and valuable comments.  This work was supported in part by 
Grants-in-Aid for Scientific Research provided by the Ministry of
Education, Science and Culture of Japan through Research Grant
No.\ 07CE2002 and No.\ 10-03687.

\newpage

Fig.\ 1. The proton chemical potential in pure neutron matter
as a function of neutron density.
The thick lines are the results calculated from Eq.\ (4) for
$C_{1}=300, 400, 600$ MeV fm$^{2}$.
The thin broken lines as marked by the Skyrme interactions
(FPS21, 1', FPS and SkM) are the results summarized by
Pethick, Ravenhall and Lorenz \cite{pethick},
and the thin solid line is the result of Sj\"oberg \cite{sj}.
The crosses denote
the values obtained by Siemens and Pandharipande
\cite{siemens}.

Fig.\ 2. The surface energy per unit area (the surface tension)
as a function of $x$, the proton fraction in the nuclear matter region.
The thick broken curves are the present results obtained from Eq.\ (6)
for $C_{1}=400$ MeV fm$^{2}$ and $C_{2}=0.01, 0.1, 1.0, 2.5, 5.0$,
the solid curve is the RBP result from their Hartree-Fock calculations
 \cite{RBP}, and the dotted curve is the BBP result \cite{BBP}.

Fig.\ 3. Zero-temperature phase diagrams on the $n_{\rm b}$ versus 
$C_2$ plane, evaluated for $C_1=300, 400, 500, 600$ MeV fm$^{2}$.

Fig.\ 4. Energy per nucleon relative to that of uniform matter
calculated for $C_{1}=400$ MeV fm$^{2}$ and $C_{2}=1.0$ 
as a function of baryon density $n_{\rm b}$.
The symbols SP, C, S, CH and SPH stand for 
sphere, cylinder, slab, cylindrical hole and spherical hole, 
respectively.

Fig.\ 5. Size of a nucleus or bubble, $r_{\rm N}$, and 
of a Wigner-Seitz cell, $r_{\rm c}$, 
calculated for $C_{1}=400$ MeV fm$^{2}$ and $C_{2}=0.01, 1.0, 2.5$ 
as a function of baryon density $n_{\rm b}$.
The symbols SP, C, S, CH and SPH stand for 
sphere, cylinder, slab, cylindrical hole and spherical hole, 
respectively.

Fig.\ 6. The neutron densities obtained for $C_{1}=300, 400, 500, 600$ 
MeV fm$^{2}$ and $C_{2}=1.0$ as functions of baryon density $n_{\rm b}$.
The lines classified by $[n(1-x)]_{\rm N}$ and $n_{\rm n}$
represent the neutron densities in the nuclear matter region 
and in the neutron gas region for the phases with nuclei, respectively.
The lines classified by $[n(1-x)]_{\rm uni}$ denote the neutron densities
in uniform nuclear matter.

Fig.\ 7. The root-mean-square displacement of a planar nucleus
at $k_{\rm B}T=0.1$ MeV, divided by the shortest distance 
between the surface of the nucleus in its equilibrium 
position and the boundary of the cell containing it.
The curves are obtained for various sets of $C_{1}$ and $C_{2}$
as a function of baryon density $n_{\rm b}$.
The thick curves lying between the two vertical lines are the results 
in the density region in which the phase with planar nuclei is 
energetically stable.

Fig.\ 8. The root-mean-square displacement of a cylindrical nucleus
at $k_{\rm B}T=0.1$ MeV, divided by the distance between 
the surface of the nucleus in its equilibrium position 
and the boundary of the cell containing it.
The curves are obtained for various sets of $C_{1}$ and $C_{2}$
as a function of baryon density $n_{\rm b}$.
The thick curves lying between the two vertical lines are the results 
in the density region in which the phase with cylindrical nuclei is 
energetically stable.

Fig.\ 9. The critical temperature $T_{\rm c}$
for the phase with planar nuclei
as a function of baryon density $n_{\rm b}$.
The thick curves lying between the two vertical lines are the results 
in the density region in which the phase with planar nuclei is
energetically stable.

Fig.\ 10. The critical temperature $T_{\rm c}$
for the phase with cylindrical nuclei
as a function of baryon density $n_{\rm b}$.
The thick curves lying between the two vertical lines are the results 
in the density region in which the phase with cylindrical nuclei is
energetically stable.

\newpage
\begin{figure}
  \begin{center}
  \vspace{-4cm}
  \psbox[height=15.5cm]{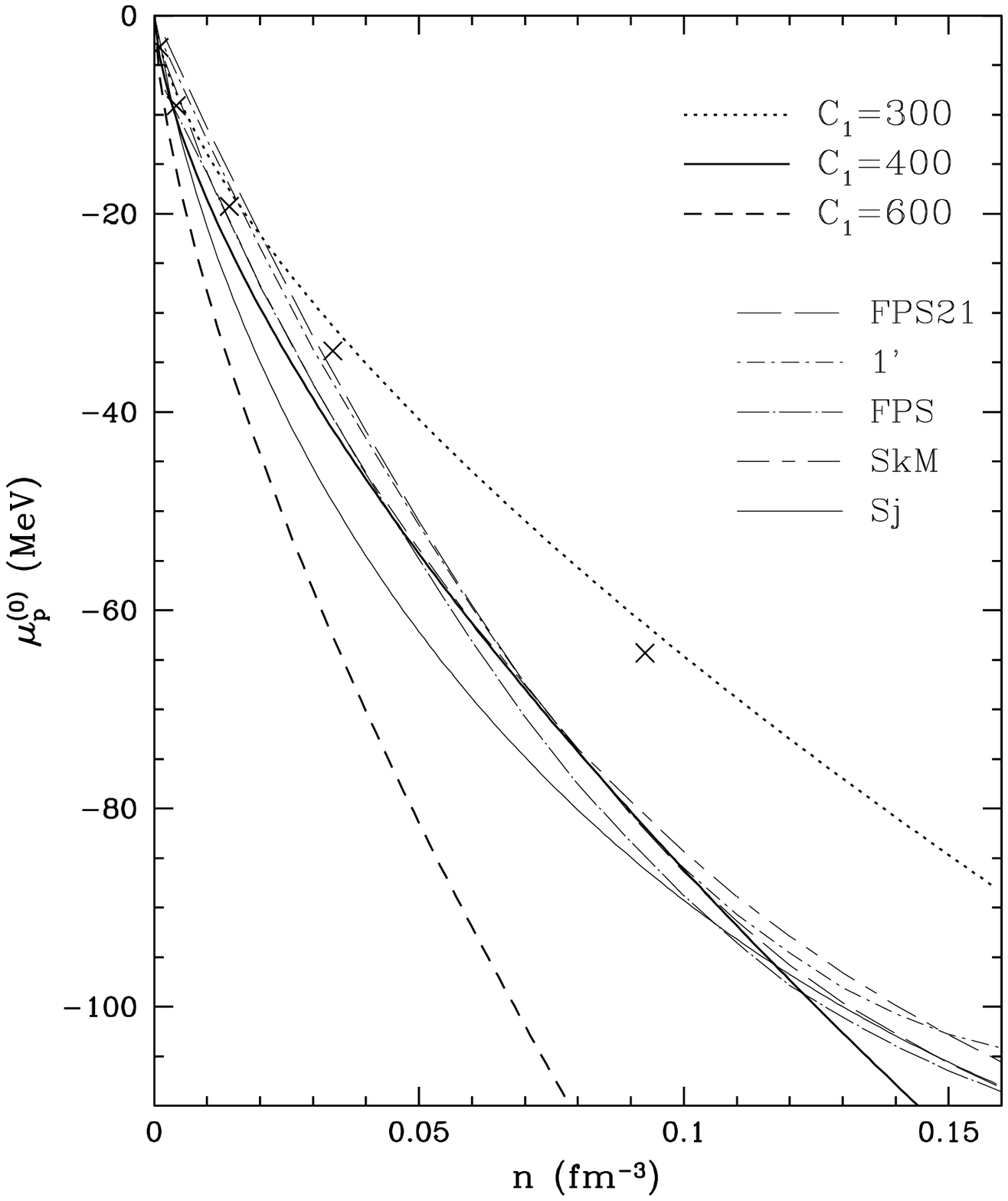}
  \end{center}
\vspace{-20pt}
\caption{}
\label{}
\end{figure}
\newpage
\begin{figure}
  \begin{center}
  \psbox[height=15.5cm]{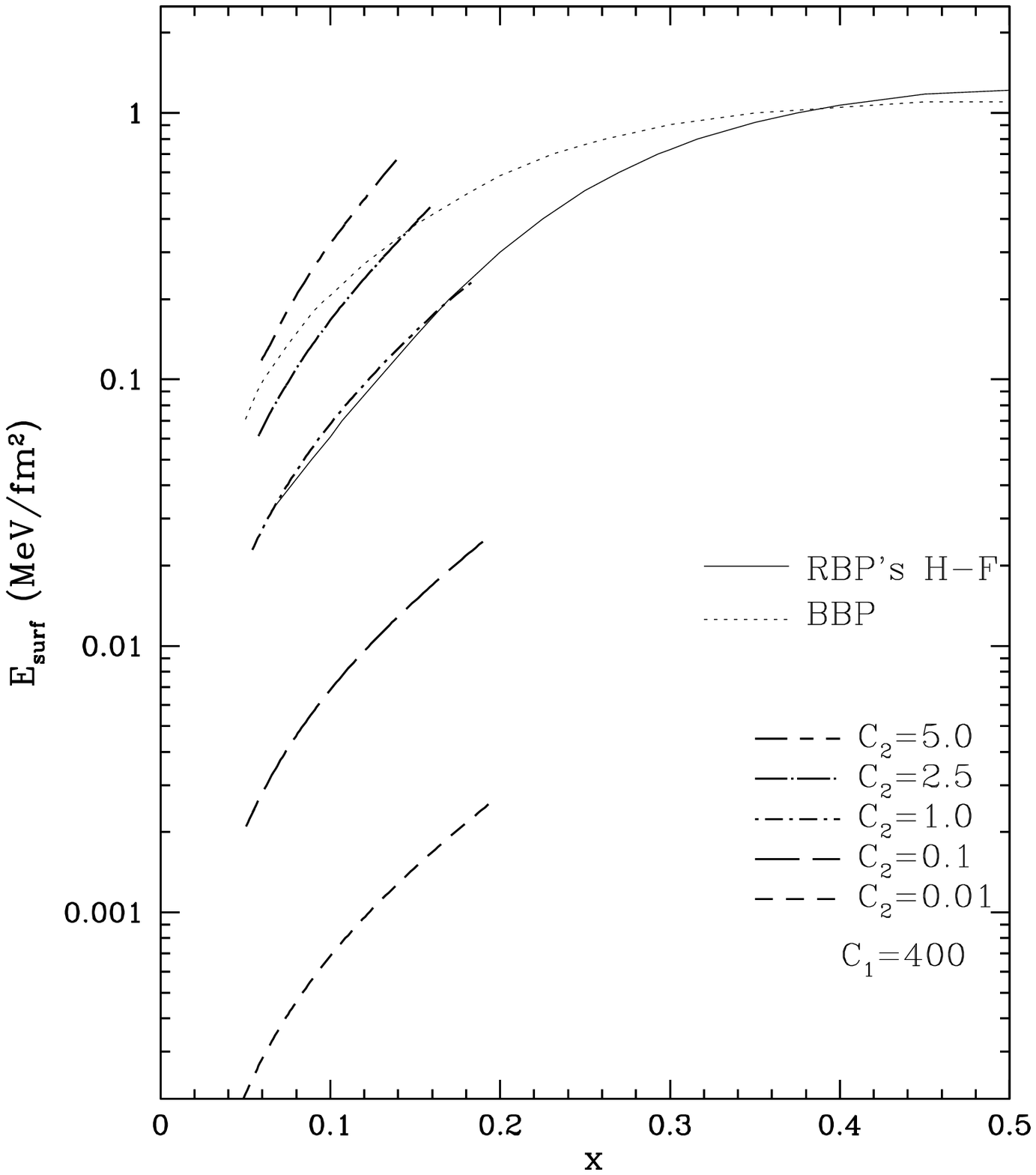}
  \end{center}
\vspace{-20pt}
\caption{}
\label{}
\end{figure}
\newpage
\begin{figure}
\hspace{2cm}
  \psbox[height=18.5cm]{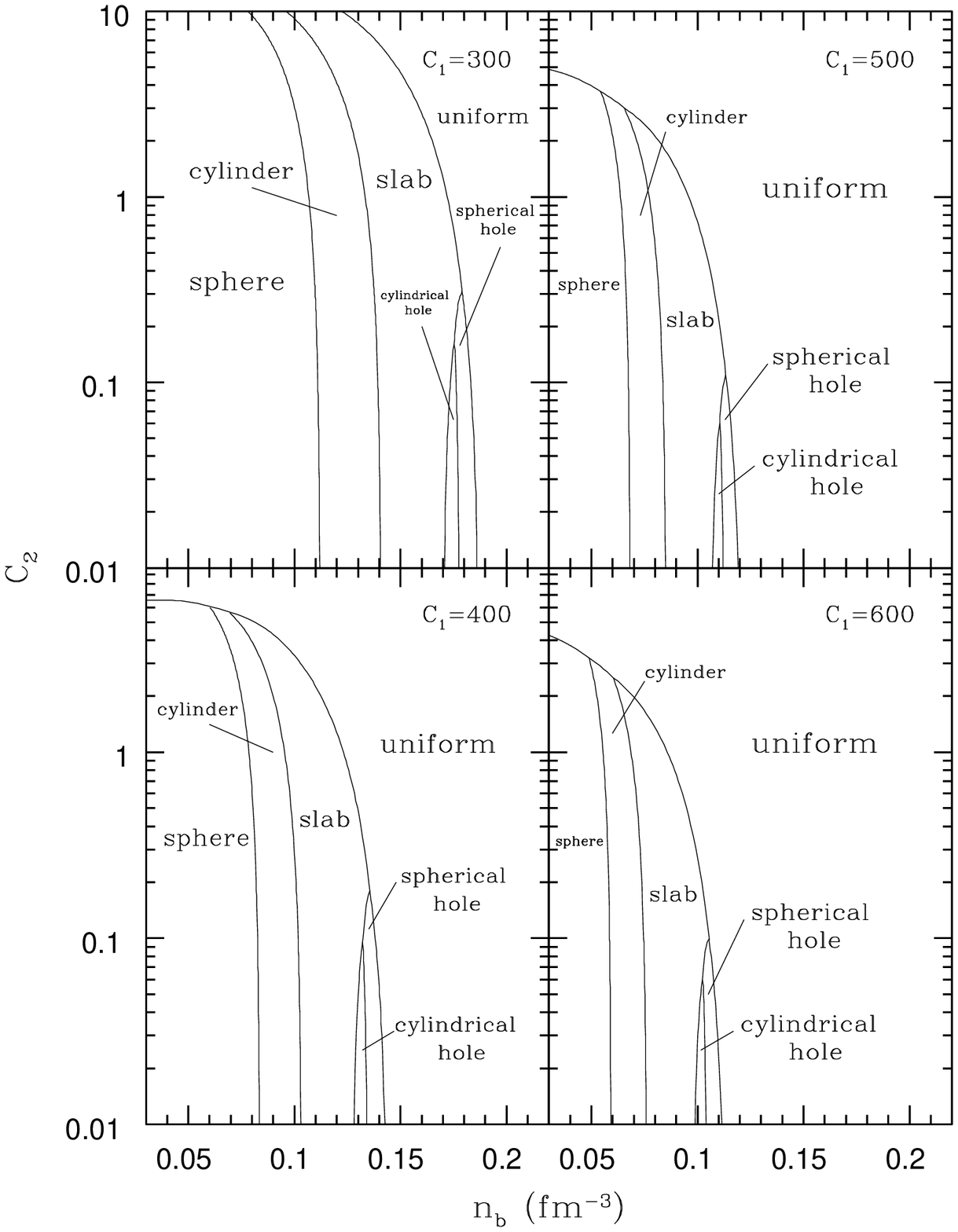}
\vspace{-20pt}
\caption{}
\label{}
\end{figure}
\newpage
\begin{figure}
  \begin{center}
  \psbox[height=18cm]{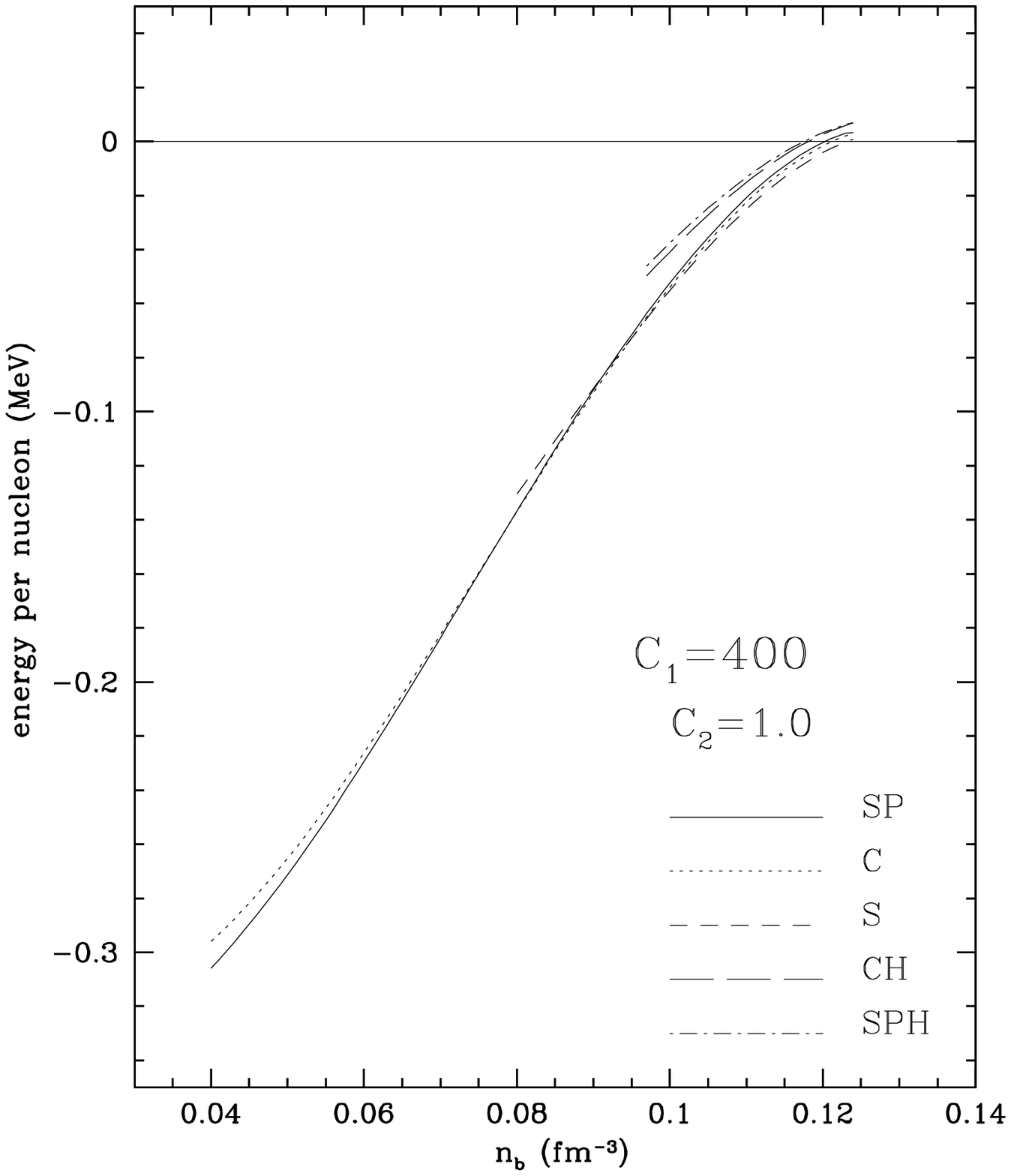}
  \end{center}
\vspace{-20pt}
\caption{}
\label{}
\end{figure}
\newpage
\begin{figure}
  \begin{center}
  \psbox[height=18cm]{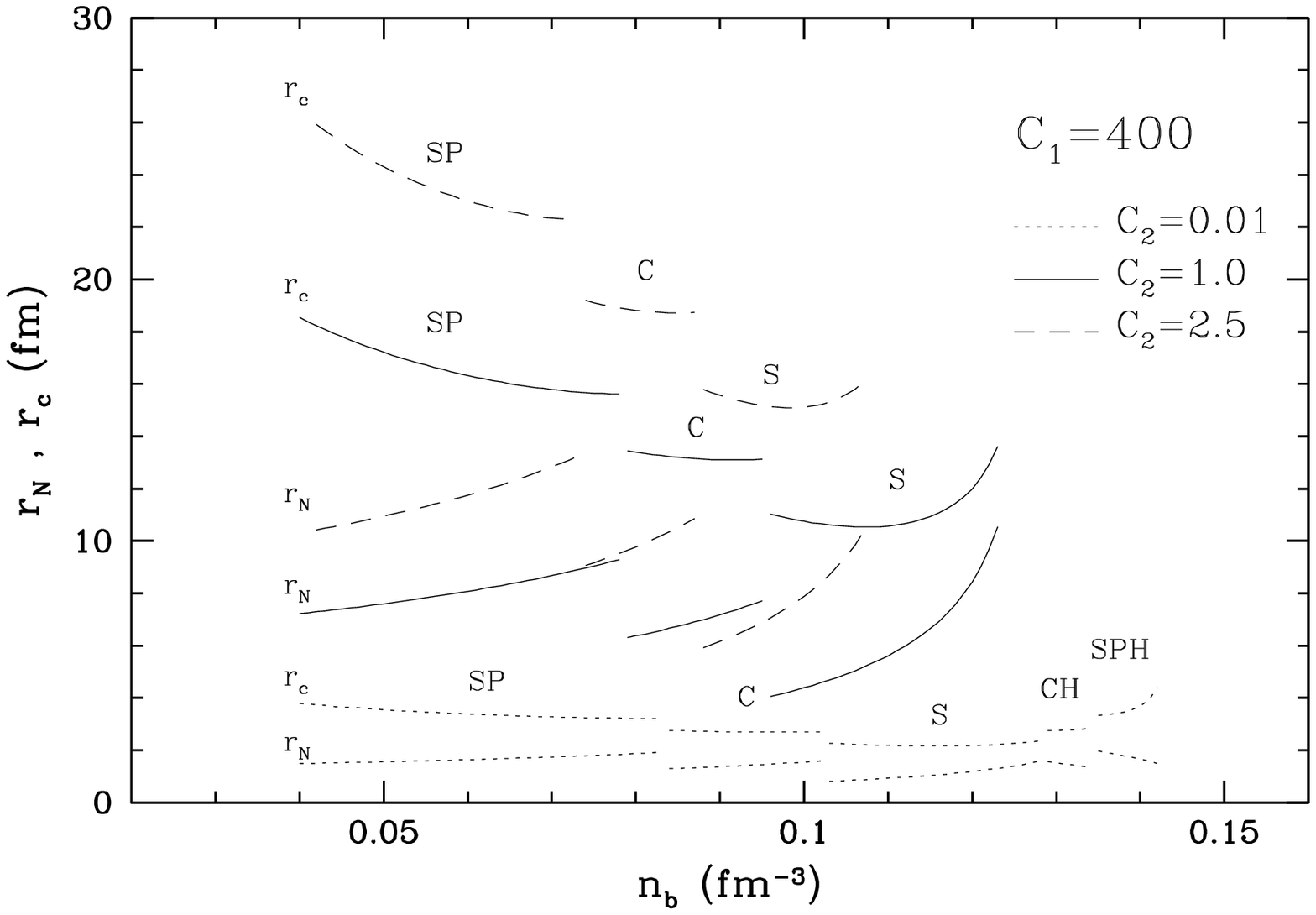}
  \end{center}
\vspace{-20pt}
\caption{}
\label{}
\end{figure}
\newpage
\begin{figure}
  \begin{center}
  \psbox[height=18cm]{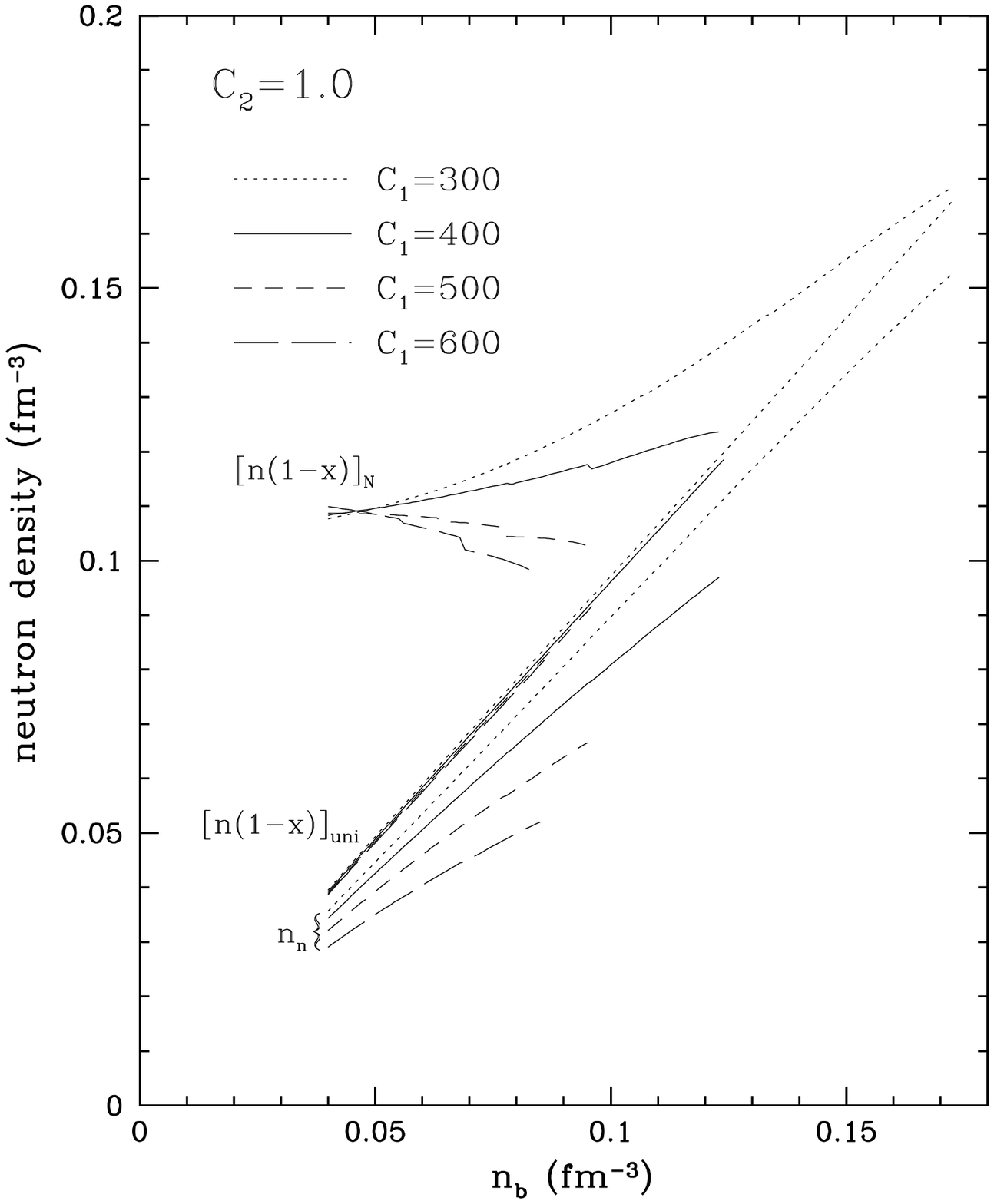}
  \end{center}
\vspace{-20pt}
\caption{}
\label{}
\end{figure}
\newpage
\begin{figure}
  \begin{center}
  \psbox[height=18cm]{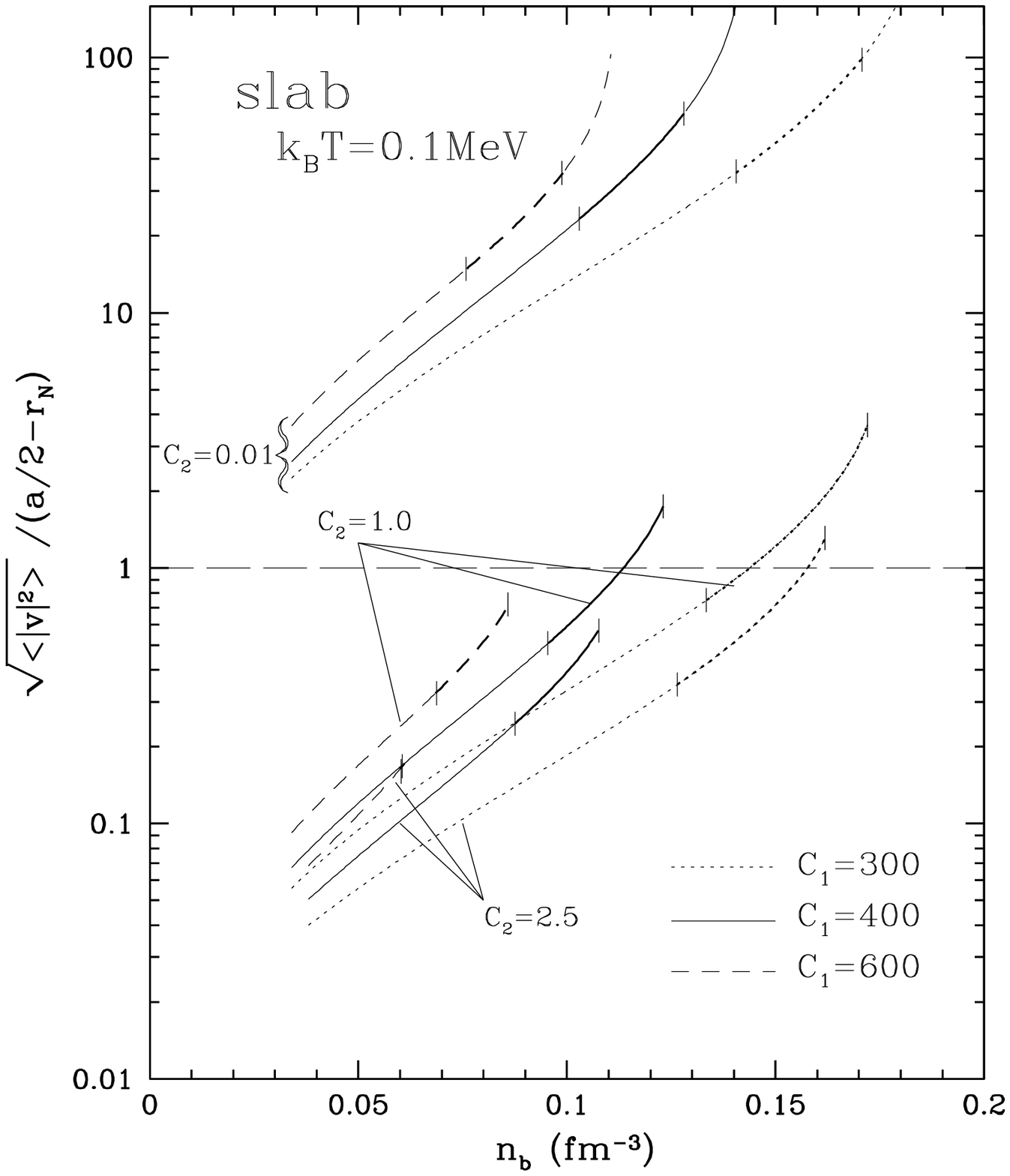}
  \end{center}
\vspace{-20pt}
\caption{}
\label{}
\end{figure}
\newpage
\begin{figure}[p]
  \begin{center}
  \psbox[height=18cm]{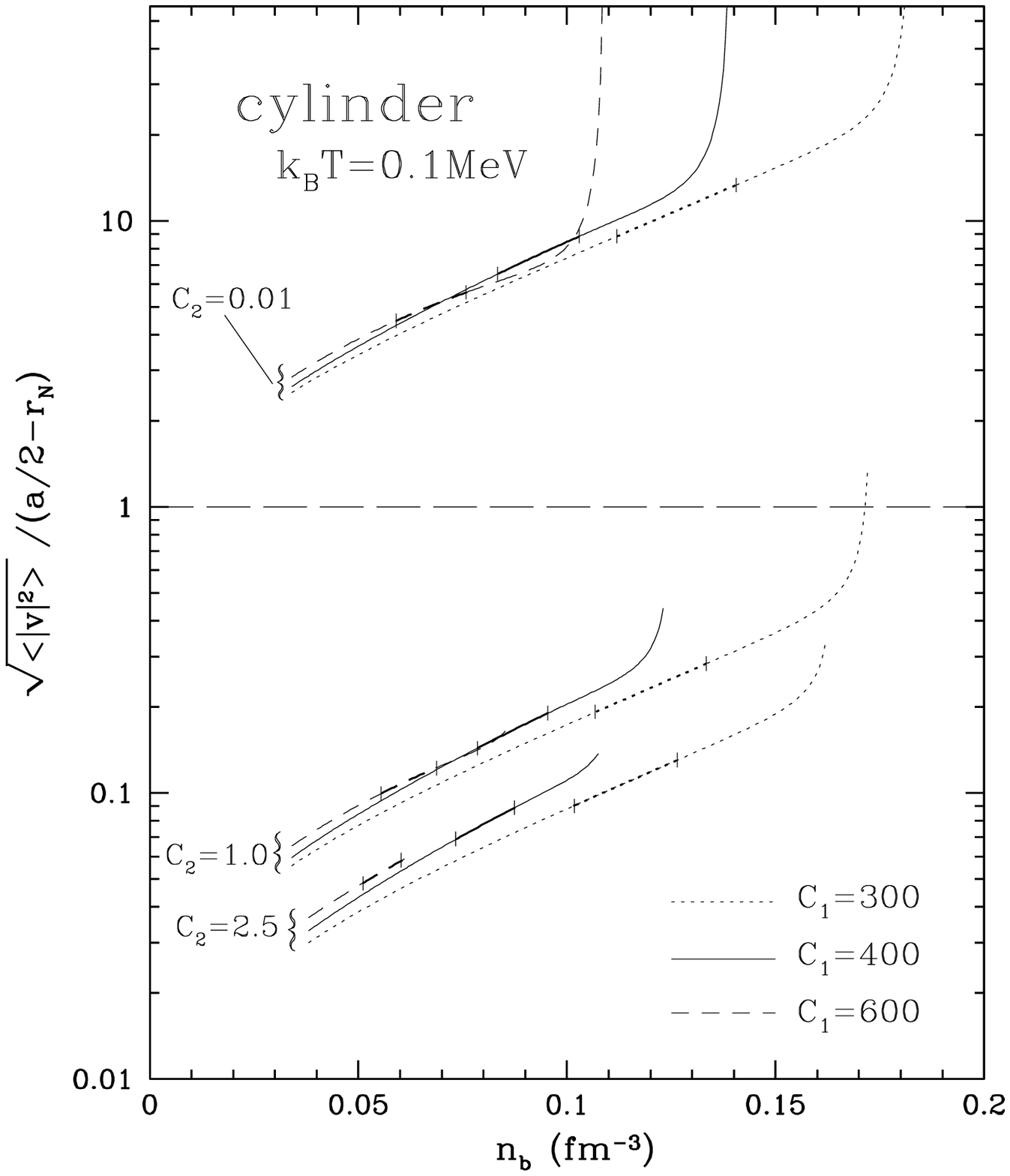}
  \end{center}
\vspace{-20pt}
\caption{}
\label{}
\end{figure}
\newpage
\begin{figure}
  \begin{center}
  \psbox[height=18cm]{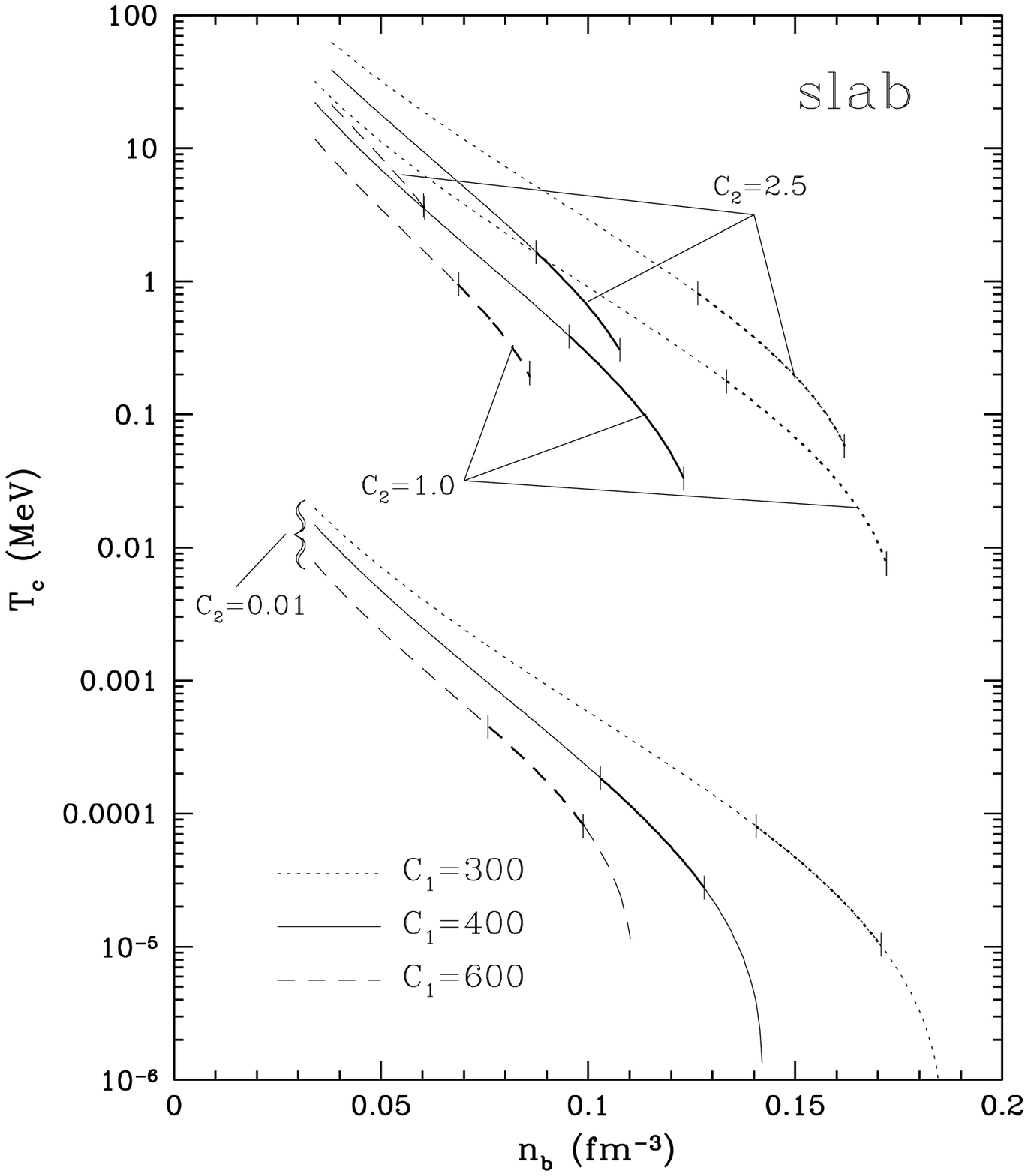}
  \end{center}
\vspace{-20pt}
\caption{}
\label{}
\end{figure}
\newpage
\begin{figure}
  \begin{center}
  \psbox[height=18cm]{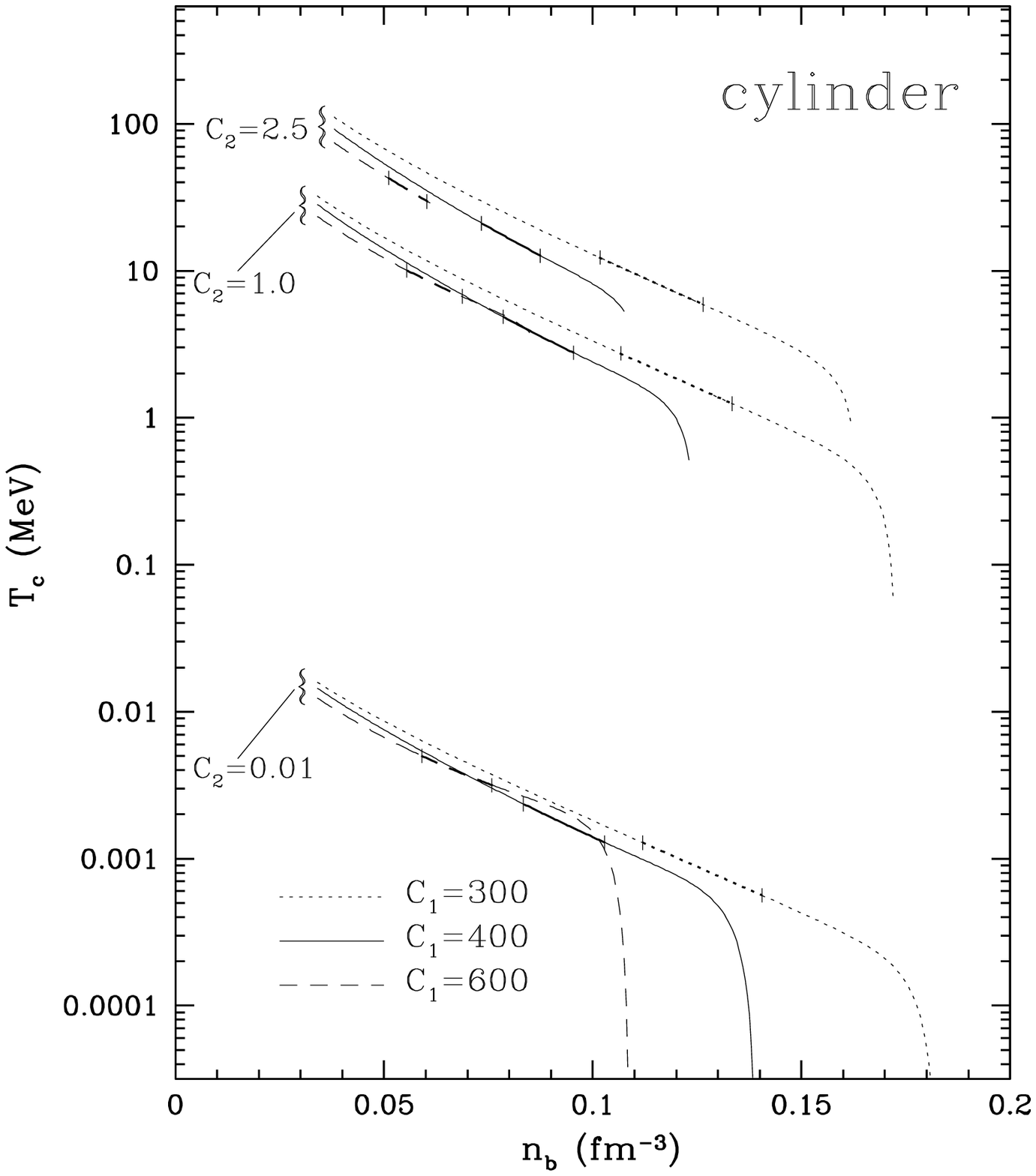}
  \end{center}
\vspace{-20pt}
\caption{}
\label{}
\end{figure}


\newpage

\begin{thebibliography}{99}
\bibitem{review} C.J. Pethick and D.G. Ravenhall,
Annu.\ Rev.\ Nucl.\ Part.\ Sci.\ 45 (1995) 429.
%
\bibitem{ravenhall} D.G. Ravenhall, C.J. Pethick and J.R. Wilson,
Phys.\ Rev.\ Lett.\ 50 (1983) 2066.
%
\bibitem{hashimoto} M. Hashimoto, H. Seki and M. Yamada,
Prog.\ Theor.\ Phys.\ 71 (1984) 320.
%
\bibitem{lorenz} C.P. Lorenz, D.G. Ravenhall and C.J. Pethick,
Phys.\ Rev.\ Lett.\ 70 (1993) 379.
%
\bibitem{oyamatsu} K. Oyamatsu,
Nucl.\ Phys.\ A561 (1993) 431.
%
\bibitem{sumiyoshi} 
K. Sumiyoshi, K. Oyamatsu and H. Toki, Nucl.\ Phys.\ A 595(1995)327.
%
\bibitem{ruderman} M. Ruderman, in {\it
Unsolved Problems in Astrophysics}, eds.\ J.N. Bahcall and J.P. Ostriker
(Princeton Univ.\ Press, Princeton, New Jersey, 1997), p.\ 281.
%
\bibitem{jones} P.B. Jones,
Mon.\ Not.\ R. Astron.\ Soc.\ 296 (1998) 217;
Mon.\ Not.\ R. Astron.\ Soc.\ 306 (1998) 327;
Phys.\ Rev.\ Lett.\ 83 (1999) 3589. 
%
\bibitem{arponen} J. Arponen,
Nucl.\ Phys.\ A191(1972)257.
%
\bibitem{BBP} G. Baym, H.A. Bethe and C.J. Pethick,
Nucl.\ Phys.\ A175 (1971)225.
%
\bibitem{potekhin} C.J. Pethick and A.Y. Potekhin,
Phys.\ Lett.\ B427 (1998) 7.
%
\bibitem{vanhorn} H.M. Van Horn, Science 252 (1991) 384.
%
%
%
%
%
%
\bibitem{pethick} C.J. Pethick, D.G. Ravenhall and C.P. Lorenz,
Nucl.\ Phys.\ A584 (1995) 675.
%
\bibitem{siemens} P.J. Siemens and V.R. Pandharipande,
Nucl.\ Phys.\ A173 (1971) 561.
%
\bibitem{sj} O. Sj\"oberg,
Nucl.\ Phys.\ A222 (1974) 161.
%
\bibitem{RBP} D.G. Ravenhall, C.D. Bennett and C.J. Pethick,
Phys.\ Rev.\ Lett.\ 28 (1972) 978.
%
\bibitem{iida} K. Iida and K. Sato,
Astrophys.\ J. 477 (1997) 294.
%
\bibitem{kolehmainen} K. Kolehmainen, M. Prakash, J.M. Lattimer and 
J.R. Treiner, 
Nucl.\ Phys.\ A439 (1985) 535.
%
\bibitem{ogasawara} R. Ogasawara and K. Sato,
Prog.\ Theor.\ Phys.\ 71 (1984) 320.
%
\bibitem{degennes} P.G. de Gennes and J. Prost,
{\it The Physics of Liquid Crystals, 2nd ed.} (Clarendon, Oxford, 1993).
%
\bibitem{chandra} S. Chandrasekhar,
{\it Liquid Crystals, 2nd ed.} (Cambridge Univ.\ Press, Cambridge, 1992).
%
\end{thebibliography}
\end{document}